\documentclass[runningheads]{llncs}
\usepackage[T1]{fontenc}
\usepackage{amsfonts}
\usepackage{amsmath}
\DeclareMathOperator*{\argmax}{argmax}
\usepackage{booktabs}
\usepackage{graphicx}
\usepackage[misc]{ifsym}
\usepackage{subfigure}
\usepackage{cleveref}
\usepackage{xcolor}
\usepackage{etoolbox}\AtBeginEnvironment{algorithm}{\tiny}
\usepackage[noend,ruled,vlined,noline]{algorithm2e}

\usepackage{mwe}
\usepackage{array}
\newcolumntype{P}[1]{>{\raggedright\arraybackslash}p{#1}}

\begin{document}

\title{
Explainable Malware Detection with Tailored Logic Explained Networks}

\titlerunning{Explainable Malware Detection with\\ Tailored Logic Explained Networks}
\author{Peter Anthony\inst{1,*} \and Francesco Giannini\inst{2} \and Michelangelo Diligenti\inst{2} \and Martin Homola\inst{1} \and Marco Gori\inst{2} \and \v{S}tefan Balogh\inst{3} \and J\'{a}n Moj\v{z}i\v{s}\inst{4}}
\institute{Department of Applied Informatics, Comenius University Bratislava, Slovakia \and
Department of Information Engineering and Mathematics, University of Siena, Italy \and
Faculty of Electrical Engineering and Information Technology Slovak University of Technology Ilkovicova 3, Slovakia \and
Institute of Informatics, Slovak Academy of Sciences, Slovakia
}
\authorrunning{P. Anthony et al.}
\maketitle              

\begin{abstract}
    Malware detection is a constant challenge in cybersecurity due to the rapid development of new attack techniques. Traditional signature-based approaches struggle to keep pace with the sheer volume of malware samples. Machine learning offers a promising solution, but faces issues of generalization to unseen samples and a lack of explanation for the instances identified as malware. 
    However, human-understandable explanations are especially important in security-critical fields, where understanding model decisions is crucial for trust and legal compliance.
    While deep learning models excel at malware detection, their black-box nature hinders explainability. Conversely, interpretable models often fall short in performance. To bridge this gap in this application domain, we propose the use of Logic Explained Networks (LENs), which are a recently proposed class of interpretable neural networks providing explanations in the form of First-Order Logic  (FOL) rules.
    This paper extends the application of LENs to the complex domain of malware detection, specifically using the large-scale EMBER dataset. In the experimental results we show that LENs achieve robustness that exceeds traditional interpretable methods and that are rivaling black-box models.   Moreover,  we introduce a tailored version of LENs that is shown to generate logic explanations with higher fidelity with respect to the model's predictions.
\end{abstract}

\section{Introduction}
Malware detection is a critical and ever-evolving task in the field of cybersecurity, where malicious actors constantly develop new techniques and exploit vulnerabilities. Software programs from a variety of firms, including Comodo, Kaspersky, Kingsoft, and Symantec, provide the primary defense against malware using a signature-based approach~\cite{MDsurvey}. Unfortunately, anti-malware providers are faced with millions of possible malware samples per year. As a result, traditional signature-based malware detection methods \cite{Borojerdi2013,Venugopal2008} may encounter significant hurdles, as malware programmers can outpace them. To keep up with the growing number of malware samples, intelligent approaches for efficient malware detection from actual and massive daily sample collections are crucial.

In this regard, machine learning methods, such as Deep Neural Networks (DNN), offer a robust solution for malware detection due to their adaptability to evolving threats, ability to recognize complex patterns, scalability for handling large datasets, feature extraction capabilities, and capacity to detect zero-day attacks. These methods may easily reduce false positives, analyze behavior, and enable continuous learning, enhancing overall detection effectiveness in the dynamic landscape of cybersecurity threats. Unfortunately, the majority of such methods are not explainable, thus limiting human trustworthiness and preventing them to be unconstrainedly used in industry, and generally in safety-critical applications.

Recently, Logic Explained Networks (LENs) \cite{CIRAVEGNA2023103822} have been proposed as an explainable-by-design class of neural networks. LENs only require their inputs to be human-understandable predicates, called concepts, and then provide explanations in terms of intelligible First-Order Logic (FOL) formulas involving such predicates. 
The main idea behind this class of models is to blend the strengths of both black and white box methods by maximising the trade-off between accuracy and interpretability of a neural model.
Roughly this is achieved as LENs determine the relevant subsets of input concepts that provide high-accuracy low-complexity explanations by using ad-hoc regularization techniques and learning objectives. 

While LENs have been successfully applied in different domains \cite{ciravegna2023learning} and to different kind of data, such as images \cite{barbiero2022entropy}, textual information \cite{jain2022extending} and graphs \cite{azzolin2023global},
their effectiveness over very large real-world datasets (such as the EMBER malware dataset~\cite{anderson2018ember})
is still unexplored. The EMBER dataset consists of 800,000 labelled samples with thousands of features, thus requiring the application of LENs to a very large dataset both in terms of features and sample size.
As a result, this paper tailors the application of explainable machine learning models to the complex landscape of malware detection, by showing that LENs can be the basis for a robust malware detection framework, which is capable of discriminating malware with performances that are competitive against state-of-the-art black-box models, and much superior than low-performing interpretable-by-design methods. While LENs have the advantage of extracting compact and meaningful explanations in the form of FOL rules that accurately match the classifier predictions, we also tailor LENs by introducing an innovative approach to enhance the fidelity of the extracted class-level explanations.

\paragraph{Contributions. } In summary, the contributions of this paper can be described as follows: (i) we investigate the effectiveness of LENs in a malware detection task, showing that LENs provide both meaningful explanations and predictive performances on-par with state-of-the-art black box models, while clearly outperforming other interpretable models, (ii) we formalize a variation of the rule extracting process from LENs, which improves over the original LENs both in terms of scalability, fidelity, complexity and predictive accuracy, (iii) we perform an in-depth analysis on the LENs' extracted rules when increasing the feature size of the inputs in terms of fidelity, complexity and accuracy.

The outline of the paper is as follows: \Cref{sec:previous_work} introduces the previous work in this area, while \Cref{sec:background} introduces LENs, the basic methods used in the paper. \Cref{sec:improvement} introduces the main methodological improvements made to LENs, to tailor the target application. \Cref{sec:experimental_design} details the experimental design and datasets, while \Cref{sec:results} reports the experimental results. Finally, \Cref{sec:conclusions} draw the final remarks and discuss the impact of this application.

\section{Previous Work}
\label{sec:previous_work}

\paragraph{Traditional Machine Learning for Malware Detection. }
Using machine learning techniques to automatically train malware detectors and learning the complex patterns behind the malware is one of the most common approaches in the literature~\cite{Millar2020,Millar2021,Ye2018}.
In particular, deep learning based on Artificial Neural Networks has emerged as a promising approach for malware detection due to its ability to learn complex patterns from large datasets and generalize to unknown samples \cite{deep4domains}.
While machine learning models have shown great potential in this domain \cite{KINKEAD2021959,HUsak8470942}, they also come with their own set of challenges like a low generalization performance, when models trained on one set of malware samples may struggle to accurately classify unseen or previously unseen samples.

Another crucial limitation of classical machine learning models is that they mainly act as black-boxes, while malware detection applications need explainability, especially when considering critical domains and the prospective legal requirements to account for algorithmic outcomes in certain use cases \cite{KINKEAD2021959}. In fact, the research community on malware detection presently acknowledges the absence of interpretability as a concern \cite{Iadarola2021,Mills2019,TonyCC,dolejvs2022interpretability}. Cybersecurity experts need to understand how models arrived at their decisions and what factors contributed to their classification, thereby also improving the overall trustworthiness of the system.

\paragraph{Interpretable AI Models  for Malware Detection. }
In contrast, interpretable machine learning models like linear regression and decision trees offer explanations for their classification results but are limited in capturing complex interactions between features \cite{Orlenko2021}. Therefore, these models prioritize interpretability over performance, often leading to inferior results compared to deep neural networks \cite{Xu2021}.
To address the trade-off between classification performance and interpretability, various techniques have been proposed to interpret the results of complex machine learning models. One of such is the permutation of feature importance methods\cite{Fumagalli_2023} introduced to interpret a wide range of machine learning models, albeit at a high computational cost. Common also is the surrogate model methods such as LIME~\cite{LIME} and SHAP~\cite{SHAP}, which use interpretable models to approximate the target model. These methods have also been explored, but their expressive ability is often not as good as the complex target model, leading to inaccurate interpretations~\cite{Nembiri2018}.

Švec et al. \cite {vsvec2024semantic}, proposed the application of various concept learning approaches to achieve an interpretable malware detection over the EMBER dataset. The authors experimented with four learning algorithms: OCEL (OWL Class Expression Learner), CELOE (Class Expression Learning for Ontology Engineering), PARCEL (Parallel Class Expression Learner), and SPARCEL (Symmetric Parallel Class Expression Learner), all implemented in DL-Learner \cite{dlleaner}, a state-of-the-art framework for supervised learning of concept descriptions in description logic \cite{dlml}. Although their experimental results show clear and understandable explanations in the form of description logic formulas, their approach is computationally expensive, and their results show low performance in discriminating malware, hence discouraging to be used in concrete applications.

\section{Background on LENs}
\label{sec:background}

Logic Explained Networks (LEN) \cite{CIRAVEGNA2023103822} represent a novel class of neural network architectures, combining the advantages of black-boxes and interpretable models through their immediate interpretability in First-Order Logic. These networks require human-understandable predicates as inputs, such as tabular data or concepts extracted from raw data by a concept classifier. The explanations provided by LENs are then expressed in terms of FOL rules involving these predicates. To ensure simplicity in explanations, LENs employ ad-hoc pruning strategies and regularization techniques during the learning process. In particular, by determining relevant subsets of the input concepts, LENs effectively provide low-complexity explanations, thereby facilitating a deeper understanding of their decision-making rationale. However, by relying on possibly complex neural processing of their inputs, LENs suitably achieve high-level performances.
Formally, a LEN $f$ can be defined as a map from $[0, 1]^{d}$-valued input concepts to $r \geq 1$ output classes, that can be used to directly classify samples and provide meaningful explanations.

\begin{figure}[t]
    \centering
    \subfigure[Local explanation for single sample]{\includegraphics[width=0.46\textwidth]{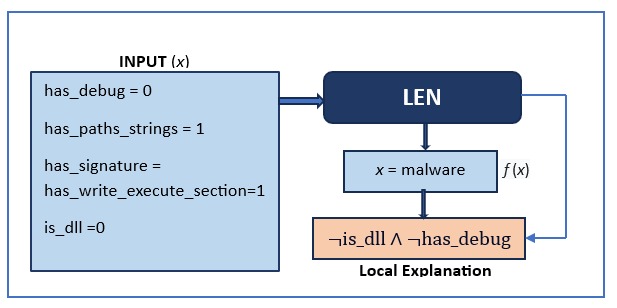}}
    \hfill
    \subfigure[Class-level explanation]{\includegraphics[width=0.5\textwidth]{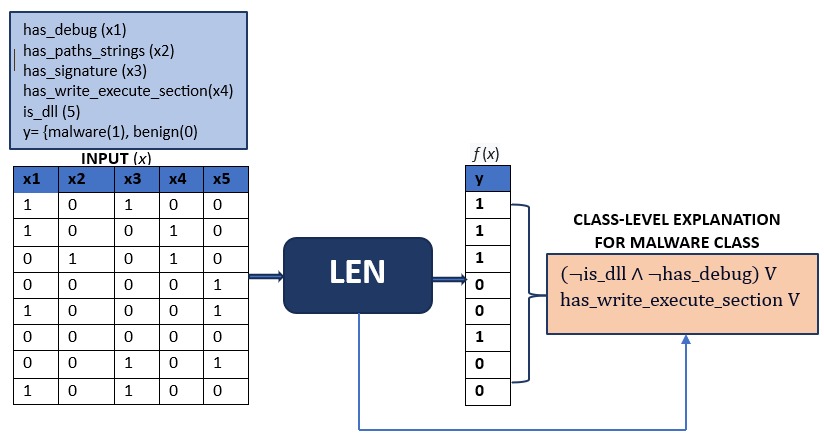}}
    \caption{Illustration of LEN local and class-level explanation for malware samples. The global explanations are obtained as a disjunction of the most important local explanations, selected using different statistical techniques.}
\end{figure}

\paragraph{Local Explanations. } For a single sample $x$,  a prediction $f_i(x) = 1$ is locally explained by the conjunction $\phi_{(l)}^{i}$
of the most relevant input features for the class $i \in \{1, . . . , r\}$. In our case, $r=2$ and
$i \in \{\textit{malware}, \textit{benign}\}$. Hence we get, 
\begin{equation}
\text{LEN Local Explanation: } \phi_l(x) = \bigwedge_{x_j \in A(i)} (\neg) x_j(x) \label{eq:local_exp}
\end{equation}
where $x_j(x)$ is a logic predicate associated with the $j$-th input feature, and $A(i)$ is the set of relevant input features for the $i$-th task. Each $x_j(x)$ can then occur as a positive $x_j(x)$ or negative $\neg x_j(x)$ literal, according to a given threshold\footnote{For instance, this threshold was fixed to 0.5 in the original paper \cite{CIRAVEGNA2023103822}. However, different values are also possible.}.

\paragraph{Global Explanation.} For global explanations, LENs consider the disjunction of the most useful local explanations:
\begin{equation}
\text{LEN Global Explanation: } \phi_g = \bigvee_{\phi_l \in B(i)} \phi_l
\label{eq:global_exp}
\end{equation}
where $B(i)$ collects the $k$-most frequent local explanations of the training set and it is computed as:
\begin{equation}
B(i) = 
\argmax_{\phi_l^{(i)} \in \Phi_l^{(i)}} \mu\left(\phi_l^{(i)}\right) 
\label{eq:kmost_exp}
\end{equation}
where we indicated with $\mu(\cdot)$ the frequency counting operator and with $\Phi_l^{(i)}$ the overall set of local explanations related to the $i$-th class. 
In the following, we will refer to  this kind of global explanation as \textbf{raw LEN explanations}.
However, this global formula could tend to be quite complex in terms of the number of used elements, which may hinder the human understanding of the explanation. For this reason, in \cite{CIRAVEGNA2023103822} the authors also proposed a top-$k$ strategy, aggregating only the top-$k$ local explanations based on their individual accuracy. Furthermore,  aggregating the local explanations (in the top-$k$) only if they improve over the validation accuracy, hence enhancing the generalization capability of the extracted rules when used to make predictions over a test set. This strategy, that we will refer to as \textbf{standard LEN explanations}, 
allows to get a more compact, accurate and understandable global explanation. However, as we will discuss in the next section, the complexity of the standard LEN explanation could still be very high in case of large-size datasets where both the number of samples and features is very high.

\section{Improving LEN's explanations}
\label{sec:improvement}

\paragraph{Motivation.}
Standard LEN explanations, even being more compact and readable than raw ones, still suffer from two main drawbacks. Firstly, they require to estimate the optimal $k$-value to use, which may be computationally costly to do. Secondly, selecting top-$k$ local explanations based on their individual accuracy, tend to favor local explanations with low precision and higher recall, and penalize local explanations with high precision and lower recall, which should be preferred to construct a robust discrimination against malware. For this reason, the extracted explanation will possibly tend to favor high false positive rate.

Jain et al.  \cite{LENp} propose an algorithm for improving the global explanation called LEN$^p$, leveraging disjunctions from all conceivable combinations of local explanations (power set). The selection of the combination with the highest accuracy serves as the global explanation.  However, this approach is intractable when the number of local explanations is very large (cf. with the paragraph below on the complexity analysis).

\begin{algorithm}[t]
\caption{Compilation of the global explanations in Tailored LENs.}
\label{fig:algo}
\DontPrintSemicolon
\SetInd{0.1em}{1em}
\SetKwFunction{FMain}{Main}
\SetKwFunction{FExp}{FilterExpl}
\SetKwFunction{EvalAcc}{EvaluateAcc}
\KwIn{$local\_expl, LEN\_Precision, \alpha$}
$Threshold_{current}\leftarrow LEN\_Precision$\;
$best\_expl\leftarrow$ \FExp{$local\_expl$, $Threshold_{current}$}\;
$Accuracy_{best}\leftarrow$ \EvalAcc{$best\_expl$}\;
\;
\While{not reached optimal accuracy}
{
$Threshold_{current}\leftarrow Threshold_{current}-\alpha$\;
$remaining\_expl\leftarrow$ \FExp{$local\_expl$, $Threshold_{current}$}\;
$Accuracy_{current}\leftarrow$ \EvalAcc{$remaining\_expl$}\;
\If{$Accuracy_{current}>Accuracy_{best}$}{
        $best\_expl\leftarrow remaining\_expl$\;
        $Accuracy_{best}\leftarrow Accuracy_{current}$}
\Else{\textbf{break}}
}
\;
$Threshold_{current}\leftarrow LEN\_Precision$\;
\While{not reached optimal accuracy}
{
$Threshold_{current}\leftarrow Threshold_{current}+\alpha$\;
$remaining\_expl\leftarrow$ \FExp{$local\_expl$, $Threshold_{current}$}\;
$Accuracy_{current}\leftarrow$ \EvalAcc{$remaining\_expl$}\;
\If{$Accuracy_{current}>Accuracy_{best}$}{
        $best\_expl\leftarrow remaining\_expl$\;
        $Accuracy_{best}\leftarrow Accuracy_{current}$}
\Else{
\textbf{break}\tcp*[f]{Optimal accuracy reached}\;
}
}
\end{algorithm}

\paragraph{Tailored-LEN explanation.}
To overcome the above discussed limitation, this paper defines an innovative approach to improve the global LEN explanation that turns out to be very useful for the malware detection application, which we call \textbf{Tailored-LEN explanation}. This methodology involves implementing a line search optimization approach for finding the optimal threshold for selecting an optimal combination. This threshold is utilized to eliminate undesirable terms, specifically those deemed as local explanations from outlier samples. 

Initially, the precision of the LEN model is computed, and the threshold is set to this precision value. Using this threshold, the best solution is obtained by aggregating a local explanation only if its precision falls within this threshold. The rationale behind this precision-driven refinement is to mitigate the risk of false positives, which could otherwise lead to a global explanation that misrepresents the model. However, a possible limitation is the fact that this approach may provide an explanation with high precision, low false positive rate, but a very low recall. To avoid this negative effect, we introduce an optimization process using a line search strategy based on the optimal threshold. 
The optimization process iteratively updates the threshold in both decreasing and increasing directions. The obtained combination performance is evaluated at each new threshold, and the best solution and accuracy are tracked. This iterative process continues until no improvement is observed in either direction. Ultimately, the optimal solution is determined by returning the simplified formula corresponding to the best solution found during the optimization process. Finally, local explanations in the optimal solution are aggregated only if they improve the validation accuracy. Algorithm \ref{fig:algo} shows the Tailored-LEN aggregation method.

\paragraph{Complexity Analysis. }
The complexity of LEN$^p$, utilizing power set combination, grows exponentially with the number of features, resulting in $O(2^n)$ complexity, where $n$ is the number of local explanations. Standard LEN, employing a Top-K strategy, scales linearly with the number of features and the selected $k$ combinations, being approximately $O(nk)$. In contrast, Tailored LEN, utilizing a line search approach with $m$ thresholds, has a complexity of approximately $O(mn)$, as it explores combinations across a range of thresholds. Hence, Tailored LEN reduces the search space by focusing on thresholds, offering a more efficient search strategy compared to LEN$^p$'s exhaustive explorationm, whereas Standard LEN provides a compromise between exhaustive search and efficiency.

\section{Experimental Analysis}
\label{sec:experimental_design}

In this section, we discuss the dataset and data preparation used, and outline the key components of our experimental design for this study.  
\subsection{Dataset and Preprocessing}
The Elastic Malware Benchmark for Empowering Researchers (EMBER), released in 2018~\cite{anderson2018ember}, is a well-known dataset of malware samples.
The EMBER dataset is the main dataset used throughout this study and provides a comprehensive collection of features extracted from Windows Portable Executable (PE) files. The dataset comprises both benign and malicious samples. It contains features from 1.1 million PE files with diverse attack types, of which $800,000$ are labelled samples ($400,000$ benign and $400,000$ malicious), and $300,000$ are unlabelled samples. We harnessed only the labelled samples for our study.

While the EMBER dataset is in JSON format, for our experimental analysis we utilized the version with derived features, as defined by Mojžiš and Kenyeres \cite{mojzis}. This dataset consists of binary features (each feature can be either true or false -- i.e. it is boolean, denoting the presence/positiveness or the absence/negativeness of each feature) to create a simplified representation. This representation is actually a variation of the ontology realized over the EMBER dataset by Švec et al. \cite{svec2022knowledgebased}.

\subsection{Feature Selections Methods}
Since the majority of  interpretable methods have severe performance limitations and also fail at providing human-readable explanations when a large feature set is available, we compared our approach against other interpretable models using the same feature selection techniques used in the original papers.
In particular, following what was done in \cite{8108975,PPR}, a decision tree-based feature selection technique was used in the experiments to identify and retain the most informative features. 
In the experiments we considered the cases with the 5, 10, 15, 20, 25, 50, 100, 200, 500, 1000, 2000 most informative features. 
On the other hand, for the comparison of LENs with state-of-the-art black-box models, like Deep Neural Networks (DNN), we used the full of set of features. 


\subsection{Evaluation Metrics}
We evaluated both the LEN model and explanations performance using standard metrics, i.e. accuracy, precision, recall, False Positive Rate and F1-score. For evaluating the explanations performance, two additional metrics were employed: Fidelity \cite{CIRAVEGNA2023103822,fidelitypapenmeier2019model} and Complexity \cite{CIRAVEGNA2023103822}.


The \textit{Fidelity} metric measures the extent to which explanations faithfully represent the inner workings of predictive models. Formally, given a data collection, a predictive model (\( \text{Model}_{\text{PM}} \)), and a model explanation (\( \text{Model}_{\text{Ex}} \)), the Fidelity(\( \text{Model}_{\text{Ex}} \)) is defined as the accuracy obtained when comparing the predictions made by \( \text{Model}_{\text{PM}} \) and the predictions derived from the explanations \( \text{Model}_{\text{Ex}} \):
\begin{equation}
\text{Fidelity}
= \frac{1}{N} \sum_{i=1}^{N} \textit{Acc}\left(\text{Model}_{\text{PM}}(x_i),\text{Model}_{\text{Ex}}(x_i)\right)
\label{eq:fidelity}
\end{equation}
where \( N \) is the number of samples in the data collection, $\textit{Acc}$ denotes the accuracy metric, and \( \text{Model}_{\text{PM}}(x_i) \) and \( \text{Model}_{\text{Ex}}(x_i) \) represent the predictions made by \( \text{Model}_{\text{PM}} \) and \( \text{Model}_{\text{Ex}} \), respectively, for the \( i \)-th sample $x_i$. This fidelity metric will serve as a crucial indicator of the trustworthiness of the explanations extracted.

The \textit{Complexity} metric counts the number of terms in the explanation as a proxy for the human understandability of the explanation.

\subsection{Case Studies}
The carried out experimental analysis has three main objectives: (i) showing that LENs performances are comparable with the ones obtained by black-box models, while also providing explanations; (ii) showing that LENs clearly outperform previous adopted interpretable machine learning models; (iii) analysing quantitatively the rules' quality with the above discussed metrics and discussing qualitatively the plausibility of the provided explanations in the malware detection application domain.

\paragraph{Experiment 1: Comparison against black-box models. } The first experiment aim at comparing with state-of-the-art black-box models, which have been applied to the full dataset of interest (i.e. the EMBER dataset) in the original papers. Hence, we used the entire 800,000 samples for this analysis (600,000 for training and 200,000 for testing) to ensure a fair comparison. In addition to having shown the LENs results over the data with the standard full amount of features, we have also reported here the results of the application of LENs after having selected the subsets of the 10, 50, 100, 200, 500, 1000 and 2000 most informative features respectively.

\paragraph{Experiment 2: Comparison against other explainable approaches. } To the author knowledge, there are two works in the literature that analyzed the EMBER dataset with the aim of interpretable malware detection: (i) the concept learning approach by Švec et al. \cite{vsvec2024semantic}, and (ii) the employment of decision-tree-based approaches by Mojžiš and Kenyeres \cite{mojzis}. For the concept learning approach, due to the complexity of their approach, only 5,000 randomly selected samples were used for a 5-fold cross-validation. Hence, in this experiment we maintained the same samples size of 5,000 and 5-fold cross-validation on the full feature set. On the other hand for the decision-tree-based approaches, their best results were obtained from the J48-ESFS tree model on 200 selected features.  Their experiment was carried out using 600,000 samples split into 80\% for training and 20\% for testing. Hence, to maintain the same set-up, we used the 200 selected features for 600,000 samples (80\% for training and 20\% for testing).

\paragraph{Experiment 3: Analysis of provided Explanations. } The third phase of our experiments focuses on the evaluation of the explanations provided by our proposed Tailored-LEN explanation method (cf. \Cref{sec:improvement}) with respect to the Raw-LEN and Standard-LEN methods, as defined in the original paper \cite{CIRAVEGNA2023103822}. To assess quantitatively the quality of the explanations, we worked with 25,000 samples, and experimented using 5, 10, 15, 20 and 30 selected features. The data is split into 75\% and 25\% training and test set respectively. The datasets were trained using the LEN model and explanations are extracted using the Raw-LEN, Standard-LEN and Tailored-LEN approaches.

In addition, to understand if the extracted rules make sense in practice, we discussed them from the perspective of malware detection applications.

\section{Results and Discussion}
\label{sec:results}

\begin{table}[t]
  \centering
  \caption{Performance comparison of LENs against Black Box Approaches.}
  \begin{tabular}{lccccccc}
    \toprule
    Model & XAI & Accuracy & Precision & Recall & FP-Rate & F1-Score \\
    \midrule
    LGBM\cite{con} & No &0.9363 & 0.9244 & 0.9504 & 0.0605 & 0.9372\\
    ANN/DNN\cite{con} & No & 0.95 & 0.96 & 0.94 & 0.0478 & 0.95 \\
    Improved DNN\cite{amol} & No & 0.9404 & 0.9014 & 0.8885 & 0.1571 & 0.8866 \\
    FFN\cite{teja} & No & - & 0.97 & 0.97 & - & 0.97 \\
    CNN\cite{teja} & No & - & 0.95 & 0.95 & - & 0.95 \\
    MalConv w/ GCG\cite{raff} & No & 0.9329 & - & - & - & - \\
     \midrule
    $LEN_{10f}$ & Yes & $0.8014$ & $0.7766$ & $0.8475$ & $0.2449$ & $0.8105$ \\
    $LEN_{50f}$ & Yes & 0.9128 & 0.09071 & 0.9202 & 0.0947 & 0.9136 \\
    $LEN_{100f}$ & Yes & 0.92074 & 0.9124 & 0.9313 & 0.0898 & 0.9217 \\
    $LEN_{200f}$ & Yes & 0.9226 & 0.9211 & 0.9248 & 0.0796 & 0.9229 \\
    $LEN_{500f}$ & Yes & 0.9225 & 0.9224 & 0.9231 & 0.0780 & 0.9227 \\
    $LEN_{1000f}$ & Yes & 0.9232 & 0.9335 & 0.9117 & 0.0652 & 0.9224 \\
    $LEN_{2000f}$ & Yes & 0.9227 & 0.9257 & 0.9196 & 0.0742 & 0.9227 \\
    {$LEN_{all\_f}$} & Yes & {0.8695}  & {0.8783}  & {0.8568} & {0.1179} & {0.8674} \\
    \bottomrule
  \end{tabular}
  \label{tab:performance_blackbox}
\end{table}

\begin{table}[th]
    \centering
    \caption{Performance comparison of LEN against explainable approaches.
    }
    \scriptsize
    \begin{tabular}{lccccc}
        \toprule
        \multicolumn{6}{c}{Comparison with concept learning approaches} \\
        \hline
        \textbf{Model} & \textbf{Accuracy} & \textbf{Precision} & \textbf{Recall} & \textbf{FP rate} & \textbf{F1} \\
        \hline
        PARCEL(0/\checkmark/\emph{X}/20) & 0.68 $\pm$ 0.01 & 0.80 $\pm$ 0.02 & 0.49 $\pm$ 0.03 & 0.12 $\pm$ 0.01 & 0.60 $\pm$ 0.03 \\
        PARCEL(0/\emph{X}/\emph{X}/5)  & 0.62 $\pm$ 0.04 & 0.90 $\pm$ 0.02 & 0.29 $\pm$ 0.09 & 0.03 $\pm$ 0.00 & 0.43 $\pm$ 0.12 \\
        PARCEL(1/\checkmark/\emph{X}/10) & 0.72 $\pm$ 0.01 & 0.71 $\pm$ 0.01 & 0.72 $\pm$ 0.02 & 0.28 $\pm$ 0.01 & 0.72 $\pm$ 0.01 \\
        PARCEL(1/\emph{X}/\emph{X}/5)  & 0.70 $\pm$ 0.02 & 0.81 $\pm$ 0.01 & 0.52 $\pm$ 0.04 & 0.12 $\pm$ 0.00 & 0.63 $\pm$ 0.04 \\
        SPARCEL(1/\checkmark/\emph{X}/20) & 0.72 $\pm$ 0.01 & 0.72 $\pm$ 0.00 & 0.73 $\pm$ 0.02 & 0.27 $\pm$ 0.00 & 0.72 $\pm$ 0.01 \\
        SPARCEL(1/\emph{X}/\emph{X}/5) & 0.64 $\pm$ 0.03 & 0.88 $\pm$ 0.04 & 0.33 $\pm$ 0.06 & 0.04 $\pm$ 0.01 & 0.48 $\pm$ 0.08 \\
        OCEL(25/\checkmark/\checkmark/5) & 0.69 $\pm$ 0.01 & 0.68 $\pm$ 0.05 & 0.74 $\pm$ 0.10 & 0.35 $\pm$ 0.12 & 0.70 $\pm$ 0.02 \\
        CELOE(25/\checkmark/\checkmark/5) & 0.68 $\pm$ 0.01 & 0.65 $\pm$ 0.03 & 0.77 $\pm$ 0.05 & 0.40 $\pm$ 0.07 & 0.70 $\pm$ 0.01 \\
    
        \textbf{LEN}& \textbf{0.87 $\pm$ 0.01} & \textbf{0.88 $\pm$ 0.02} & \textbf{0.87 $\pm$ 0.02} & \textbf{0.13 $\pm$ 0.02} & \textbf{0.88 $\pm$ 0.01} \\
        \hline
    \bottomrule
    \end{tabular}
    \label{tab:performance_explanable}
\end{table}

\begin{figure}[t]
    \centering
    
    \subfigure[Accuracy, highest is the best. LEN vs 3 decision tree models, five different feature sizes
 counts.]{\includegraphics[width=0.8\textwidth]{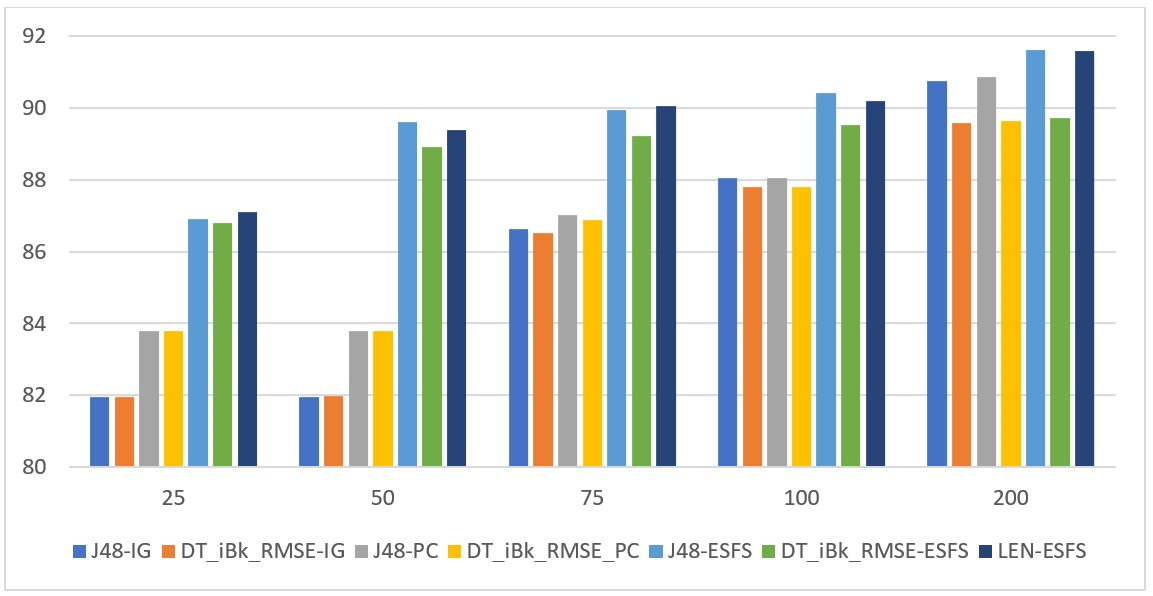}}
    \hfill

    \subfigure[False positive rate, lowest is the best. LEN vs Decision tree models for 25 features]{\includegraphics[width=0.8\textwidth]{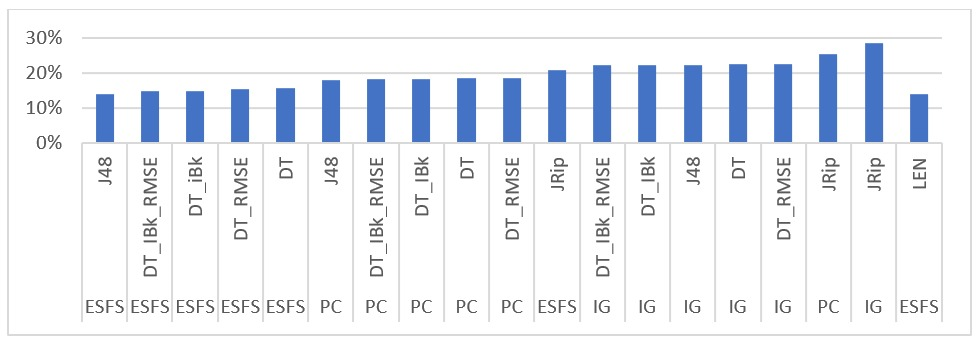}}
    \hfill

    \caption{Plots comparing the performance of the LENs in terms of (a) accuracy and (b) FP-Rate on the EMBER dataset.}
    \label{fig:len_vs_dt}
\end{figure}

In the following, we report and discuss the main results of the experiments, showing the strengths of LENs for the goal of developing an effective malware detection solution. 

\paragraph{Results of comparing against black-boxes.}
As shown in Table~\ref{tab:performance_blackbox}, LENs achieve an accuracy of at least 92.07\%, and F1-score of 92.17\% when using a minimum of 100 features (except for using the full features), which demonstrates the robustness of this model in the intricate landscape of malware detection.  The table also shows a comparison of different LEN models LEN$_{nf}$ for different considered feature sizes $n$. 
While training the model using the full set of features resulted in weaker performance compared to using a subset of features, it's worth noting that the feature binarization process applied to the features might have contributed to this discrepancy. Binarization can lead to a significant increase in feature dimensionality. According to potentially introducing noise into the dataset. This noise could affect the model's ability to generalize effectively and thus result in inferior performance when utilizing the full set of features.

LEN's competitiveness with black-box models is evident based on its close proximity across all metrics to the best deep-learning black-box models with a percentage difference of less than 5\% across most metrics.
LENs can already get competitive results using a relatively small set of features, but unlike other interpretable methods they retain high generalization capabilities even when using very large feature sizes.
However, the distinguishing factor of the proposed method lies in LEN's interpretability, setting it apart from the traditional black box. 

\paragraph{Results of comparing against interpretable models.}

In comparison to interpretable models, as indicated in Table~\ref{tab:performance_explanable}, LEN showcases a far superior performance, surpassing the performance of the concept learning approaches by a big margin across all metrics. 
The results show how LENs are a promising solution for real-world deployment, aligning with the growing demand for both transparency and robustness in malware detection applications.

Moreover, we compared LENs with standard decision tree models.
The results shown in \Cref{fig:len_vs_dt} prove that LENs perform significantly better than the majority of decision tree models, while the results are comparable with the J48-ESFS model.

\begin{table}[th]
  \centering
  \caption{On the left we have some examples of local explanations for a detected malware. On the right side it is reported a justification of the plausibility of the rule according to domain expertise in the field of malware detection.}
  \begin{tabular}{P{5.8cm}|P{5.8cm}} 
    \hline
    \textbf{Explanation} & \textbf{Cybersecurity expert remarks} \\
    \hline
   \texttt{has\_section\_high\_entropy} $\land$ $\neg$\texttt{is\_dll} $\land$ $\neg$\texttt{has\_debug} &  It points to packed malware that is not .dll and  has not  debug symbols  enabled (which is a typical malware behaviour). \\
   \hline
    \texttt{has\_write\_execute\_section} $\land$ $\neg$\texttt{has\_debug} & Malware can typically  use a section with write and execute permission for self injection.\\
    \hline
    \texttt{has\_section\_high\_entropy} $\land$ $\neg$\texttt{has\_signature} & This also point to packed malware and malware usually has no signature\\ 
   \hline
    \texttt{has\_section\_high\_entropy} $\land$ \texttt{sect\_text\_write\_execute\_section} & It point to packed (encrypted) code \\
    \hline
  \end{tabular}
  \label{tab:rule_examples}
\end{table}

\paragraph{Analysis of the Tailored-LENs explanations.}

Table~\ref{tab:rule_examples} shows some local explanations provided by LENs together with some remarks provided by a cybersecurity expert. The expert highlighted that all the explanations indicated meaningful reasons for the sample being a malware, and that it is impressive to be able to have this level of insight into the workings of an ML-based model that was able to process the full EMBER dataset. At the same time, all explanations were more general and abstract compared to those derived by concept learning on a fractional dataset \cite{vsvec2024semantic}.

\begin{figure}[t]
    \centering
    
    \subfigure[Comparison based on accuracy]{\includegraphics[width=0.45\textwidth]{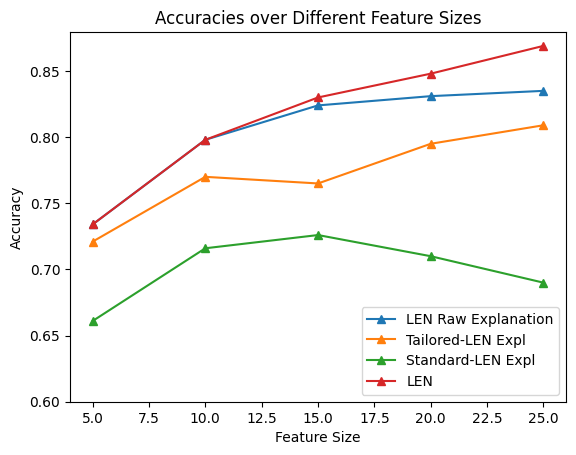}}
    \hfill
    \subfigure[Comparison based on FP-Rate]{\includegraphics[width=0.45\textwidth]{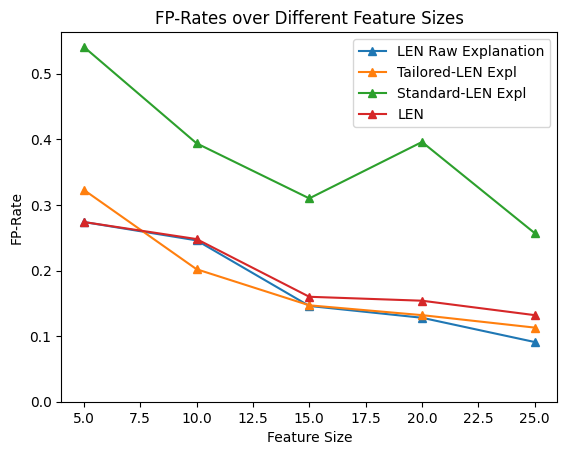}}

    \vspace{0.5em} 

    \subfigure[Comparison based on fidelity]{\includegraphics[width=0.45\textwidth]{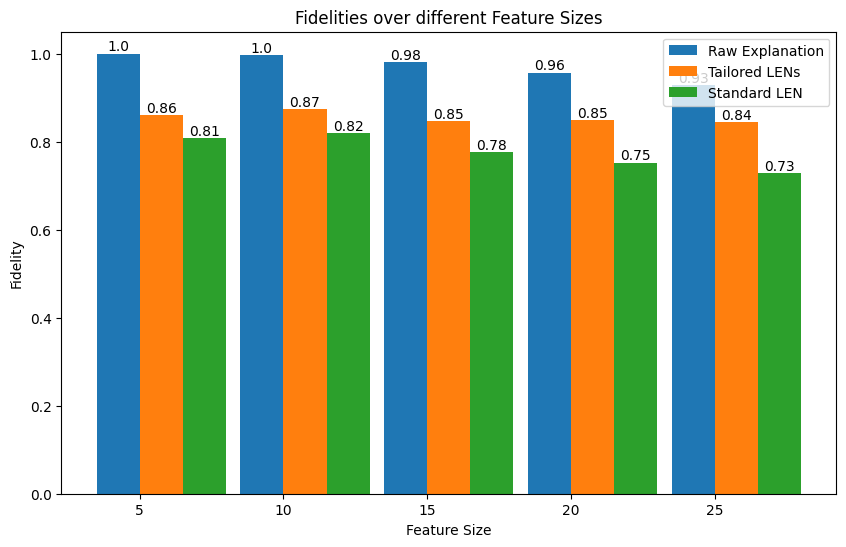}}
    \hfill
    \subfigure[Comparison based on complexity of explanations]{\includegraphics[width=0.45\textwidth]{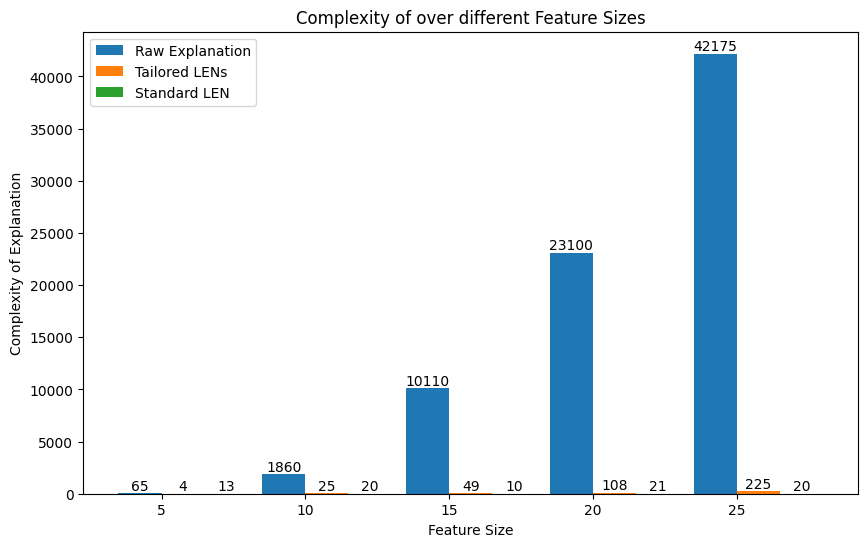}}

    \caption{Plots comparing the performance of the explanations of the different LENs in terms of (a) accuracy, (b) FP-Rate, (c) Fidelity and (d) Complexity, over different feature size on the EMBER dataset.}
    \label{fig:len_comparison}
\end{figure}

Figure~\ref{fig:len_comparison} compares the performances of the different LENs over the EMBER dataset for different feature selection sizes. Tailored LEN explanations significantly improve over Standard LENs for all tested feature sizes, while also providing better fidelity and lower complexity. Raw LEN explanations are the best in pure terms of predicting performance, but they are plagued by a very high explanation complexity, which completely hinders the human understandably of what the model is doing.
On the other hand, Standard-LEN explanations have a complexity that is similar to Tailored-LENs but they can not match neither the fidelity nor the accuracy of Tailored-LENs.

\section{Conclusions}
\label{sec:conclusions}


This paper studies the application of Logic Explained Networks to the task of malware detection. The experiments have shown that LENs can achieve performance levels similar to complex black-box models, while still being explainable, and outperforming interpretable machine learning competitors.
Moreover, this paper proposed a new algorithm to get a global explanation from LENs, which preserves the predictive accuracy of LENs, while providing explanations that have both high fidelity and low complexity. Supported by these evidences, we think that these results pave the way toward the application of explainable methods to large-scale malware detection tasks.



%
%
%
\bibliographystyle{splncs04}
\bibliography{ref}


\end{document}